\documentclass[conference]{IEEEtran}

\usepackage{ascmac}
\usepackage{graphicx}
\usepackage{cite}
\usepackage{listings}
\usepackage{paralist}
\usepackage{nidanfloat}
\usepackage{fancybox}
\usepackage{float}
\usepackage{amsmath}
\usepackage{amsfonts}
\usepackage[font=footnotesize]{caption}
\usepackage[font=footnotesize]{subcaption}
\usepackage[hyphens]{url}
\usepackage{pifont}
\usepackage{multirow}
\usepackage{comment}
\usepackage{tabularx}
\usepackage{textcomp}
\usepackage[dvipdfmx]{color}
\usepackage{fixltx2e}
\usepackage[keeplastbox]{flushend}

\usepackage{epsfig}
\usepackage{setspace}

\usepackage{hanging}
\usepackage[flushmargin,bottom]{footmisc}

\def\BibTeX{{\rm B\kern-.05em{\sc i\kern-.025em b}\kern-.08em
    T\kern-.1667em\lower.7ex\hbox{E}\kern-.125emX}}

\begin{document}

\title{Experimental Analysis of Communication Relaying Delay in Low-Energy Ad-hoc Networks}

\author{\IEEEauthorblockN{
Taichi Miya\IEEEauthorrefmark{1},
Kohta Ohshima\IEEEauthorrefmark{2},
Yoshiaki Kitaguchi\IEEEauthorrefmark{1} and
Katsunori Yamaoka\IEEEauthorrefmark{1}}
\IEEEauthorblockA{\IEEEauthorrefmark{1}Tokyo Institute of Technology,
Ookayama 2--12--1--S3--68,
Meguro-ku, Tokyo, 152--8552 Japan\\
E-mail: miya@net.ict.e.titech.ac.jp, kitaguchi@gsic.titech.ac.jp, yamaoka@ict.e.titech.ac.jp}
\IEEEauthorblockA{\IEEEauthorrefmark{2}Tokyo University of Marine Science and Technology,
Etchujima 2--1--6, Koto-ku,
135--8533 Japan\\
E-mail: kxoh@kaiyodai.ac.jp}}
\maketitle

\begin{abstract}
In recent years, more and more applications use ad-hoc networks for local M2M communications,
but in some cases such as when using WSNs, the software processing delay induced by packets relaying may not be negligible.
In this paper, we planned and carried out a delay measurement experiment using Raspberry Pi Zero W.
The results demonstrated that, in low-energy ad-hoc networks, processing delay of the application is always too large to ignore;
it is at least ten times greater than the kernel routing and corresponds to 30\% of the transmission delay.
Furthermore, if the task is CPU-intensive, such as packet encryption,
the processing delay can be greater than the transmission delay and its behavior is represented by a simple linear model.
Our findings indicate that the key factor for achieving QoS in ad-hoc networks is an appropriate node-to-node load balancing
that takes into account the CPU performance and the amount of traffic passing through each node.
\end{abstract}


\IEEEpeerreviewmaketitle

\section{Introduction and Related Works} \label{sec:introduction}
An ad-hoc network is a self-organizing network that operates independently of pre-existing infrastructures
such as wired backbone networks or wireless base stations by having each node inside the network behave as a repeater.
It is a kind of temporary network that is not intended for long-term operation.
Every node of an ad-hoc network needs to be tolerant of dynamic topology changes
and have the ability to organize the network autonomously and cooperatively.

Because of these specific characteristics, since the 1990s, ad-hoc networks have played an important role
as a mean for instant communication in environments where the network infrastructure is weak or does not exist,
such as developing countries, disaster areas, and battle fields.
However, in recent years, the ad-hoc network is also a hot topic in urban areas
where the broadband mobile communication systems are well developed and always available.
More and more applications use ad-hoc networks for local M2M communications,
especially in key technologies that are expected to play a vital role in future society,
such as intelligent transportation systems (ITS) supporting autonomous car driving,
cyber-physical systems (CPS) like smart grids,
wireless sensor networks (WSN), and applications like the IoT platform.

These days, communication entities are shifting from humans to things;
the network infrastructures tend to require a more strict delay guarantee, and the ad-hoc network is no exception.
There have been many prior studies about delay-aware communication in the field of ad-hoc networks\cite{myself,unified,hello,multipath}.
Most of these focus on the link delay and only a few consider both node and link delays\cite{myself,unified}.
However, in some situations where the power consumption is severely limited (e.g., with WSN),
the communication relaying cost of small devices with low-power processors
may not be negligible for the end-to-end delay of each communication.

It is necessary to discuss, on the basis of actual data measured on wireless ad-hoc networks,
how much the link and node delays account for the end-to-end delay.
In the field of wired networks, there have been many studies reporting measurement experiments of packet processing delay
as well as various proposals for performance improvement\cite{Papagiannaki,Dx,coffeebreak,Angrisani,Beifuss,Salehin}.
In addition, the best practice of QoS measurement has been discussed in the IETF\cite{ippm}.
In the past, measurement experiments on ASIC routers have been carried out
for the purpose of benchmarking routers working on ISP backbones\cite{Papagiannaki,Dx,coffeebreak};
in contrast, since the software router has emerged as a hot topic in the last few years,
recent studies mainly concentrate on the bottleneck analysis of the Linux kernel's network stack\cite{Angrisani,Salehin,Beifuss}.
There has also been a study focusing on the processing delay caused by the low-power processor
assuming interconnection among small robots\cite{robot}.
However, as far as we know, no similar measurement exists in the field of wireless ad-hoc networks.
Therefore, many processing delay models have been considered so far,
e.g., simple linear approximation \cite{model1} or queueing model-based nonlinear approximation \cite{model2},
but it is hard to determine which one is the most reasonable for wireless ad-hoc networks.

In this work, we analyze the communication delay in an ad-hoc network through a practical experiment using Raspberry Pi Zero W.
We assume an energy-limited ad-hoc network composed of small devices with low-power processors.
Our goal is to support the design of QoS algorithms on ad-hoc networks
by clarifying the impact of software packet processing on the end-to-end delay
and presenting a general delay model to which the measured delay can be adapted.
This is an essential task for future ad-hoc networks and their related technologies.

First, we briefly describe the structure of the Linux kernel network stack in Sect.~\ref{sec:kernel}.
We explain the details of our measurement experiment in Sects.~\ref{sec:design} and ~\ref{sec:method},
and report the results in Sect.~\ref{sec:result}.
We conclude in Sect.~\ref{sec:conclusion} with a brief summary and mention of future work.

\section{Network Stack of Linux Kernel} \label{sec:kernel}
In this section, we present a brief description of the Linux kernel's standard network stack
from the viewpoints of the packet receiving and sending sequences.

\begin{figure}[!t]
  \centering
  \includegraphics[height=53mm]{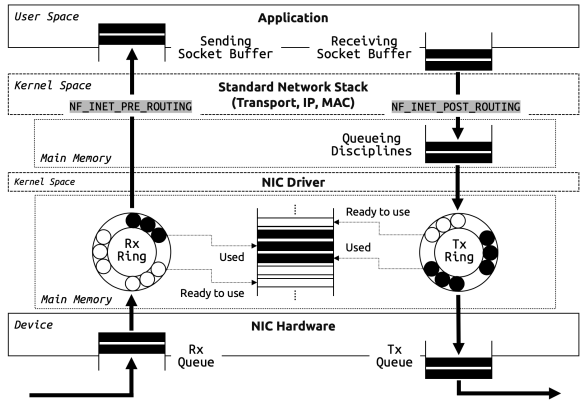}
  \caption{Packet queueing in the Linux network stack.}
  \label{fig:networkstack}
  \vspace{0mm}
\end{figure}

\subsection{Generic Packet Receiving Sequence}
Figure~\ref{fig:networkstack} shows the flow of packets in the network stack from the perspective of packet queueing.

First, as the preparation for receiving packets, the NIC driver allocates memory resources in RAM that can store a few packets,
and has packet descriptors (Rx descriptors) hold these addresses.
The Rx ring buffer is a descriptor ring located in RAM, and the driver notifies the NIC of the head and tail addresses of the ring.
The NIC then fetches some unused descriptors by direct memory access (DMA) and waits for the packets to arrive.
The workflow after the packet arrival is as follows.
As a side note, the below sequence is a receiving mechanism called new API (NAPI) supported in Linux kernel 2.6 or later.

\vspace{4mm}
\begin{enumerate}
  \renewcommand{\labelenumi}{\roman{enumi})}
  \item Once a packet arrives, NIC writes the packet out as an sk\_buff structure to RAM with DMA,
  referring to the Rx descriptors cached beforehand, and issues a HardIRQ after the completion.
  \item The IRQ handler receiving HardIRQ pushes it by \texttt{napi\_schedule()}
  to the poll\_list of a specific CPU core and then issues SoftIRQ so as to get the CPU out of the interrupt context.
  \item The soft IRQ scheduler receiving SoftIRQ calls the interrupt handler \texttt{net\_rx\_action()} at the best timing.
  \item \texttt{net\_rx\_action()} calls \texttt{poll()}, which is implemented in not the kernel but the driver, for each poll\_list.
  \item \texttt{poll()} fetches sk\_buff referring to the ring indirectly and pushes it to the application on the upper layer.
  At this time, packet data is transferred from RAM to RAM; that is, the data is copied from the memory in the kernel space
  to the receiving socket buffer in the user space by \texttt{memcpy()}.
  Repeat this memory copy until the poll\_list becomes empty.
  \item The application takes the payload from the socket buffer by calling \texttt{recv()}.
  This operation is asynchronous with the above workflows in the kernel space.
  The packet receiving sequence is completed when all the payloads have been retrieved.
\end{enumerate}

\subsection{Generic Packet Sending Sequence}
In the packet sending sequence, all the packets basically follow the reverse path of the receiving sequence,
but they are stored in a buffer called QDisc before being written to the Tx ring buffer (Fig.~\ref{fig:networkstack}).

The ring buffer is a simple FIFO queue that treats all arriving packets equally.
This design simplifies the implementation of the NIC driver and allows it to process packets fast.
QDisc corresponds to the abstraction of the traffic queue in the Linux kernel
and makes it possible to achieve a more complicated queueing strategy than FIFO
without modifying the existing codes of the kernel network stack or drivers.

QDisc supports many queueing strategies; by default, it runs in \texttt{pfifo\_fast} mode.
If the packet addition fails due to a lack of free space in QDisc, the packet is pushed back to the upper layer socket buffer.

\section{Experimental Design} \label{sec:design}
As discussed in Sect.~\ref{sec:introduction}, the goal of this study is to evaluate the impact of software packet processing,
induced by packet relaying, to the end-to-end delay, on the basis of an actual measurement assuming an ad-hoc network
consisting of small devices with low-power processors.
Figure~\ref{fig:ad-hoc} shows our experimental environment, whose details are described in Sect.~\ref{sec:method}.

\subsection{Definitions of Delays in Network}
We define the classification of communication delays as below.
Both \textit{processing delay} and \textit{queueing delay} correspond to the application delay in a broad sense.

\begin{itemize}
  \item \textbf{End-to-end delay:} Total of node delays and link delays
  \item \textbf{Node delay:} Sum of \textit{processing delay}, \textit{queueing delays},
  and any processing delays occurring in the network stack
  \item \textbf{Link delay:} Sum of all other delays occurring under the NIC driver,
  such as queueing delay of the NIC internal buffer, transmission delay and propagation delay at the communication medium
  \item \textbf{\textit{Processing delay}:} Processing delay for packet relaying generated by the application in user space
  \item \textbf{\textit{Queueing delay}:} Queueing delay at the socket buffer
\end{itemize}

\subsection{Proxy: How to Relay Packets} \label{sec:proc}
The proxy node (Fig.~\ref{fig:ad-hoc}) relays packets with the three methods below,
and we evaluate the effect of each in terms of the end-to-end delay.
By comparing the results of \textit{OLSR} and \textit{AT},
we can clarify the delay caused by packets passing through the network stack.

\begin{itemize}
  \item \textbf{Kernel routing \textit{(OLSR)}:}
  Proxy relays packets by kernel routing based on the OLSR routing table.
  In this case, the relaying process is completed in kernel space because all packets are wrapped in L3 of the network stack.
  Accordingly, both \textit{processing delay} and \textit{queueing delay} defined above become zero,
  and node delay is purely equal to the processing delay on the network stack in the kernel space.
  \item \textbf{Address translation \textit{(AT)}:}
  Proxy works as a TCP/UDP proxy, and all packets are raised to the application running in the user space.
  The application simply relays packets by switching sockets, which is equivalent to a fixed-length header translation.
  \item \textbf{Encryption \textit{(Enc)}:}
  Proxy works as a TCP/UDP proxy.
  Besides \textit{AT}, the application also encrypts payloads using AES 128-bit in CTR mode
  so that the relaying load depends on the payload size.
\end{itemize}

\subsection{Measurement Conditions}
For each relaying method, we conduct measurements with variations of the following conditions.
We express all the results as multiple percentile values in order to remove delay spikes.
Because the experiment takes several days, we record the RSSI of the ad-hoc network including five surrounding channels.

\begin{itemize}
  \item \textbf{Payload size}
  \item \textbf{Packets per second (pps)}
  \item \textbf{Additional CPU load \textit{(stress)}}
\end{itemize}

\section{Measurement Methods} \label{sec:method}
In this section, we explain the technical details of the experimental environment and measurement programs.

\subsection{Building OLSR Ad-hoc Network}
We use three Raspberry Pi Zero Ws (see Table~\ref{tbl:specsheet} for the hardware specs).
The Linux distributions installed on the Raspberry Pis are Raspbian and the kernel version is 4.19.97+.

We use OLSR (RFC3626), which is a proactive routing protocol, and adopt \texttt{olsrd} as its actual implementation.
Since all three of the nodes are location fixed, even if we used a reactive routing protocol like AODV instead of OLSR,
only the periodic Hello in OLSR will change the periodic RREQ induced by the route cache expiring;
that is, in this experiment, whether the protocol is proactive or reactive does not have a significant impact on the final results.

\begin{figure}[t]
  \centering
  \includegraphics[width=80mm]{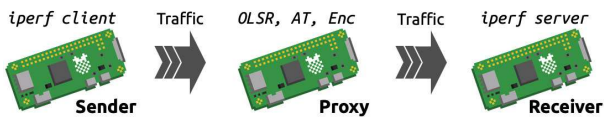}
  \caption{Three Raspberry Pis connected logically inline.}
  \label{fig:ad-hoc}
  \vspace{-4mm}
\end{figure}

\begin{table}[b]
  \begin{center}
    \small
    \caption{Hardware specs of Raspberry Pi Zero W.}
    \label{tbl:specsheet}
    \begin{tabularx}{85mm}{ll}
      \hline
      SoC & Broadcom BCM2835 \\
      CPU & ARM1176JZF-S (ARMv6) 1core 1GHz \\
      RAM & LPDDR2 SDRAM 512MB \\
      Wi-Fi & IEEE 802.11b,g,n 2.4GHz \\
      Power & 150mA (0.75W) \\
      Size & 65mm x 30mm (9g) \\
      \hline
    \end{tabularx}
  \end{center}
\end{table}

The ad-hoc network uses channel 9 (2.452 GHz) of IEEE 802.11n, transmission power is fixed to --31 dBm, and bandwidth is 20 MHz.
As WPA (TKIP) and WPA2 (CCMP) do not support ad-hoc mode, the network is not encrypted.

\begin{figure}[H]
  \centering
  \includegraphics[width=65mm]{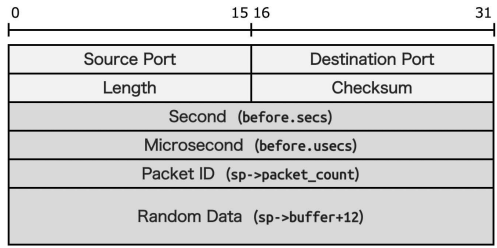}
  \caption{Format of UDP probe generated by iperf.}
  \label{fig:iperf}
\end{figure}

Although the three nodes can configure an OLSR mesh, as they are located physically close to each other,
we have the sender/receiver drop OLSR Hello from the receiver/sender as well as the ARP response by Netfilter
so that the network topology becomes a logically inline single-hop network, as show in Fig.~\ref{fig:ad-hoc}.

\subsection{Preparation of Traffic Generator}
We use \texttt{iperf} as a traffic generator
and measure the UDP performance as it transmits packets from sender to receiver via proxy.
The iperf embeds two timestamps and a packet ID in the first 12 bytes of the UDP data section (Fig.~\ref{fig:iperf}),
and the following measurement programs we implement use this ID to identify each packet.
Random data are generated when iperf starts getting entropy from \texttt{/dev/urandom},
and the same series is embedded in all packets.

\subsection{Implementation of the Kernel Module for Measurement}
We create a loadable kernel module using Netfilter and measure the \textit{queueing delay} in receiving and sending UDP socket buffers.
The workflow is summarized as follows:
the module hooks up the received packets with \texttt{NF\_INET\_PRE\_ROUTING} and the sent packets with \texttt{NF\_INET\_POST\_ROUTING}
(Fig.~\ref{fig:networkstack}),
retrieves the packet IDs iperf marked by indirectly referencing the sk\_buff structure,
and then writes them out to the kernel ring buffer via \texttt{printk()} with a timestamp obtained by \texttt{ktime\_get()}.

\subsection{Implementation of the Proxy}
The proxy program is the application running in the user space.
It creates \texttt{AF\_INET} sockets between sender and proxy as well as between proxy and receiver
and then translates IP addresses and port numbers by switching sockets.
Furthermore, it records the timestamps obtained by \texttt{clock\_gettime()} immediately after calling \texttt{recv()} and \texttt{sendto()},
and encrypts every payload data protecting the first 12 bytes of metadata marked by iperf so as not to be rewritten.
The above refers to the UDP proxy; the TCP proxy we prepare simply using \texttt{socat}.

\subsection{Additional CPU Load at Proxy}
We execute a dummy process whose CPU utilization rate is limited by \texttt{cpulimit} as a controlled noise of the user space
in order to investigate and clarify its impact on the node delay.

\section{Results and Discussion} \label{sec:result}
We performed the delay measurement experiments under the conditions shown in Table~\ref{tbl:params}
using the methods described in the previous section.
Due to the space constraints, we omit the results of the preliminary experiment.
Note that all experiments were carried out at the author's home;
due to the Japanese government's declaration of the COVID-19 State of Emergency,
we have had to stick to the ``Stay home'' initiative unless absolutely necessary.

\subsection{Time Variation of Received Signal Strength Indicator}
\begin{figure*}[!t]
  \begin{minipage}{0.5\hsize}
    \centering
    \includegraphics[width=90mm]{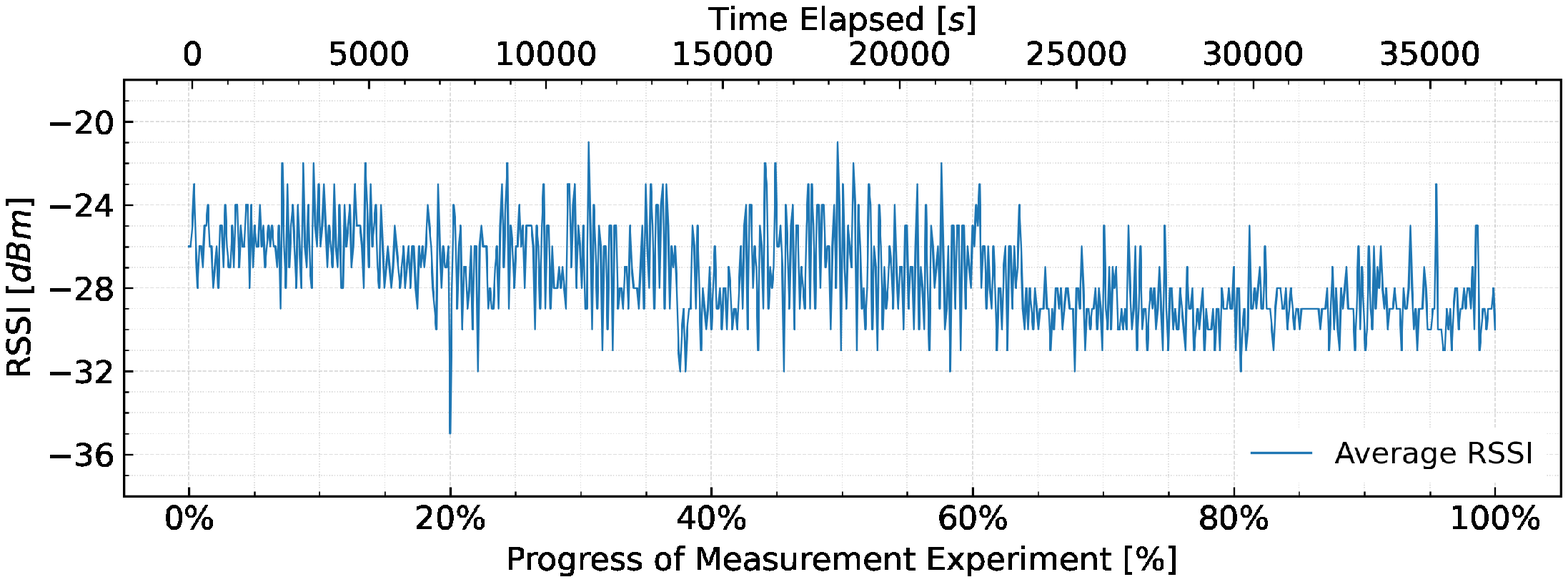}
    \subcaption{\textbf{RSSI:} Average scores for sender, proxy, and receiver.}
    \label{fig:rssi}
  \end{minipage}
  \begin{minipage}{0.5\hsize}
    \centering
    \includegraphics[width=90mm]{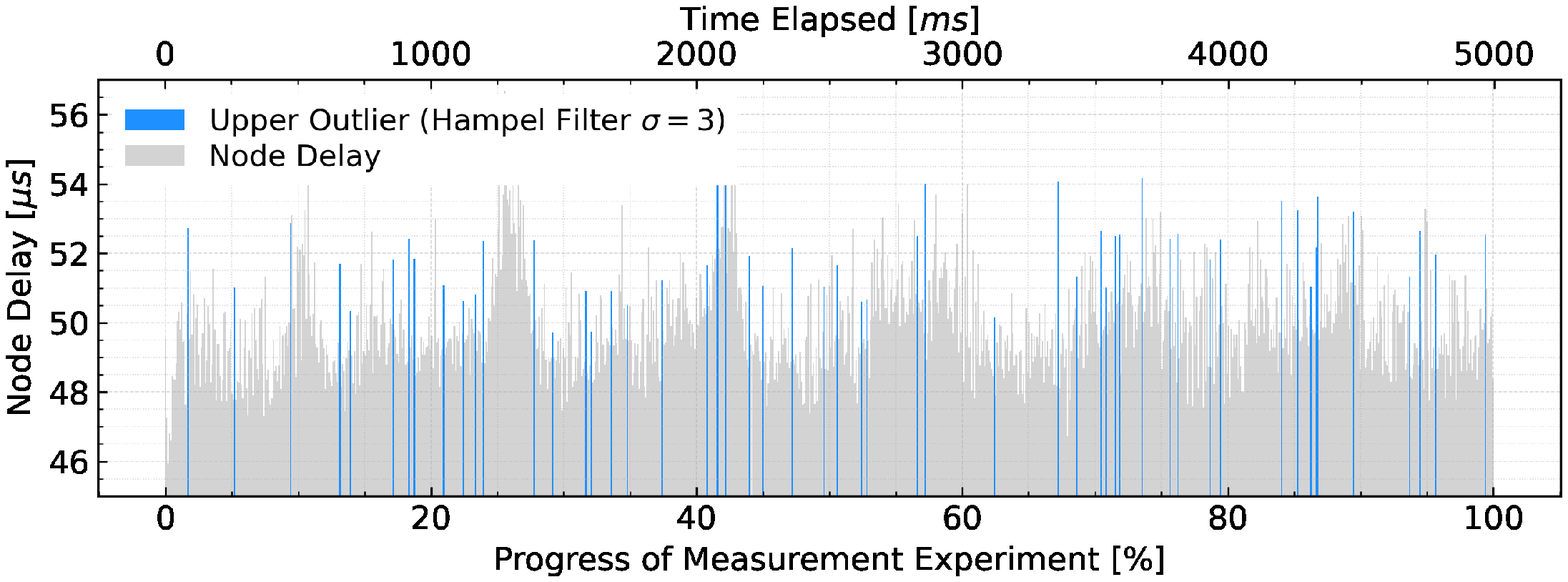}
    \subcaption{\textbf{Node delay (\textit{OLSR}):} 1000 bytes, 200 pps, 0\% \textit{stress}}
    \label{fig:timeline-olsr}
  \end{minipage}
  \begin{minipage}{0.5\hsize}
    \vspace{3mm}
    \centering
    \includegraphics[width=90mm]{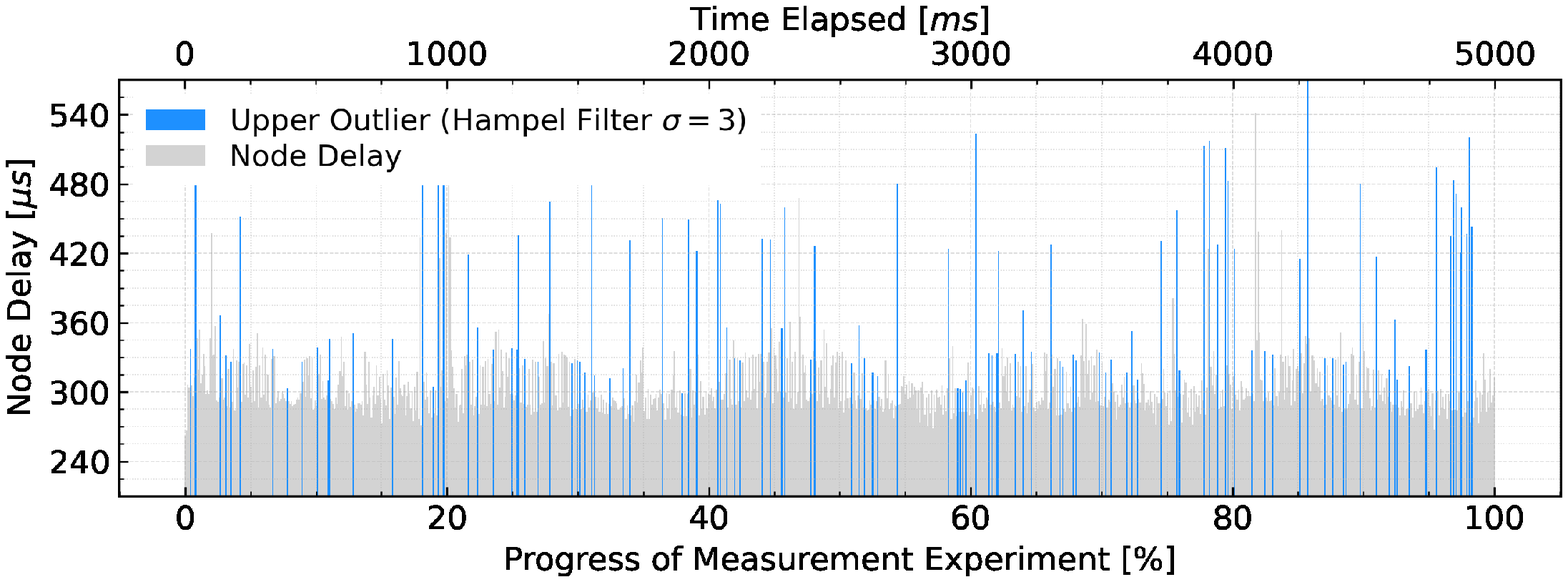}
    \subcaption{\textbf{Node delay (\textit{AT}):} 1000 bytes, 200 pps, 0\% \textit{stress}}
    \label{fig:timeline-at}
  \end{minipage}
  \begin{minipage}{0.5\hsize}
    \vspace{3mm}
    \centering
    \includegraphics[width=90mm]{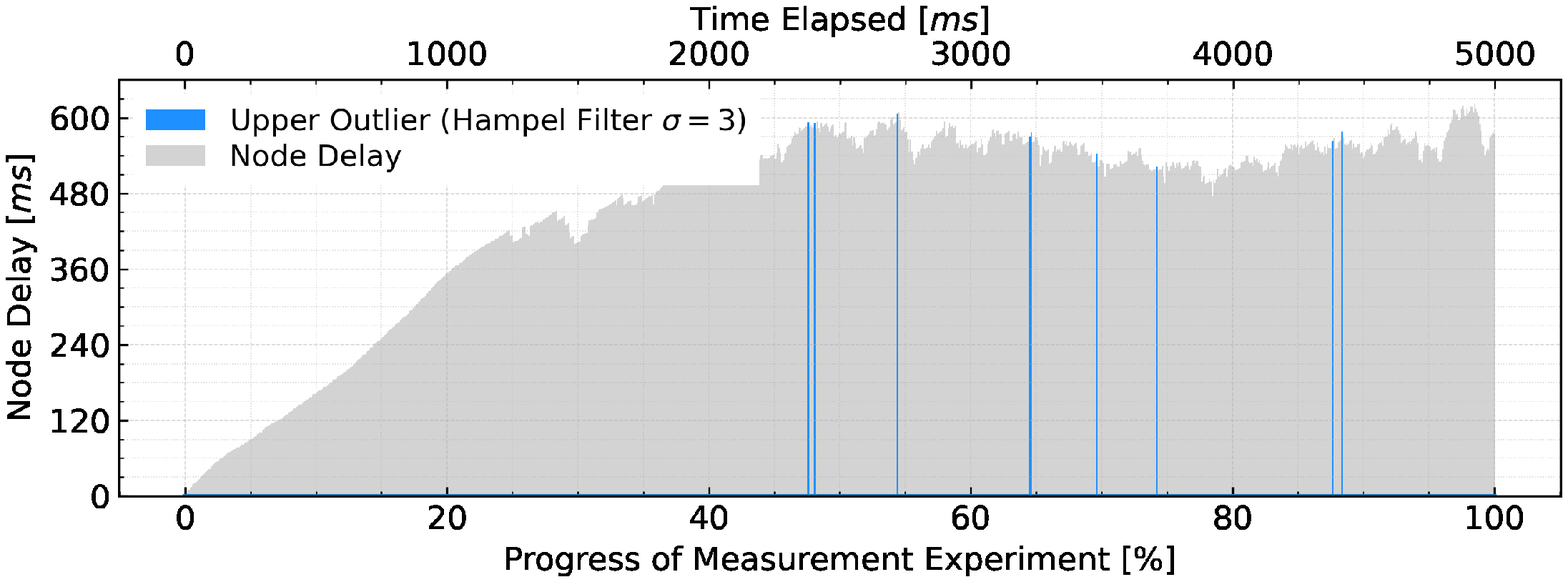}
    \subcaption{\textbf{Node delay (\textit{Enc}):} 1000 bytes, 200 pps, 0\% \textit{stress}}
    \label{fig:timeline-enc}
  \end{minipage}
  \caption{\textbf{Timelines:} Average time variation of RSSI and 100th percentile node delays during the measurement.}
  \vspace{-3mm}
\end{figure*}

The experiment was divided into nine measurements.
Figure~\ref{fig:rssi} shows the time variation of RSSI during a measurement.
We were unable to obtain SNRs owing to the specifications of the Wi-Fi driver, and thus the noise floors were unknown,
but the ESSIDs observed in the five surrounding channels were all less than --80 dBm.
The RSSI variabilities were also within the range that did not affect the modulation and coding scheme (MCS)\cite{mcs};
therefore, it appears that the link quality was sufficiently high throughout all measurements.

\subsection{Time Variation of Node Delay and Coffee-break Effect}
Figures~\ref{fig:timeline-olsr}, \ref{fig:timeline-at}, and \ref{fig:timeline-enc} shows the average time variations of node delay,
which were the results under the condition of 1000 bytes, 200 pps, and 0\% \textit{stress}.
The blue highlighted bars indicate upper outliers (delay spikes) detected with a Hampel filter ($\sigma = 3$).
There were 53 outliers in \textit{OLSR}, 115 in \textit{AT}, and 9 in \textit{Enc}.

In general, when the CPU receives periodic interrupts (e.g., routing updates, SNMP requests, GCs of RAM),
packet forwarding is paused temporarily so that the periodic delay spikes can be observed in the end-to-end delay.
This phenomenon is called the ``coffee-break effect''\cite{coffeebreak}
and has been mentioned in several references\cite{Beifuss,Papagiannaki,Angrisani}.

For this experiment, as seen in the results of \textit{AT} (Fig.~\ref{fig:timeline-at}), in the low-energy ad-hoc networks,
it is evident that the CPU-robbing by other processes like coffee-break had a significant impact on the communication delay.
Incidentally, there were fewer spikes under both \textit{\textbf{1)} OLSR} and \textit{\textbf{2)} Enc} than under \textit{AT}.
\textbf{\textit{1)}} Since the packet forwarding was completed within the kernel space,
node delay was less susceptible to applications running in the user space.
\textbf{\textit{2)}} Since the payload encryption was overwhelmingly CPU-intensive,
the influence of other applications was hidden and difficult to observe from the node delay.

\begin{figure*}[!t]
  \begin{minipage}{0.33\hsize}
    \centering
    \includegraphics[width=60mm]{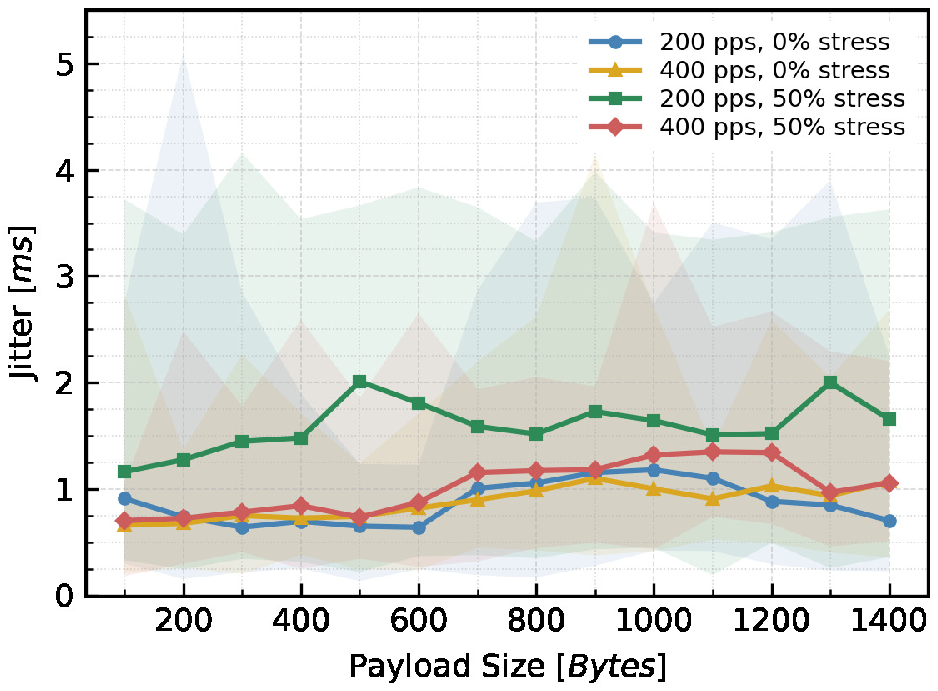}
    \subcaption{\textbf{Jitter \textit{(AT)}}}
    \label{fig:jitter-at}
  \end{minipage}
  \begin{minipage}{0.33\hsize}
    \centering
    \includegraphics[width=60mm]{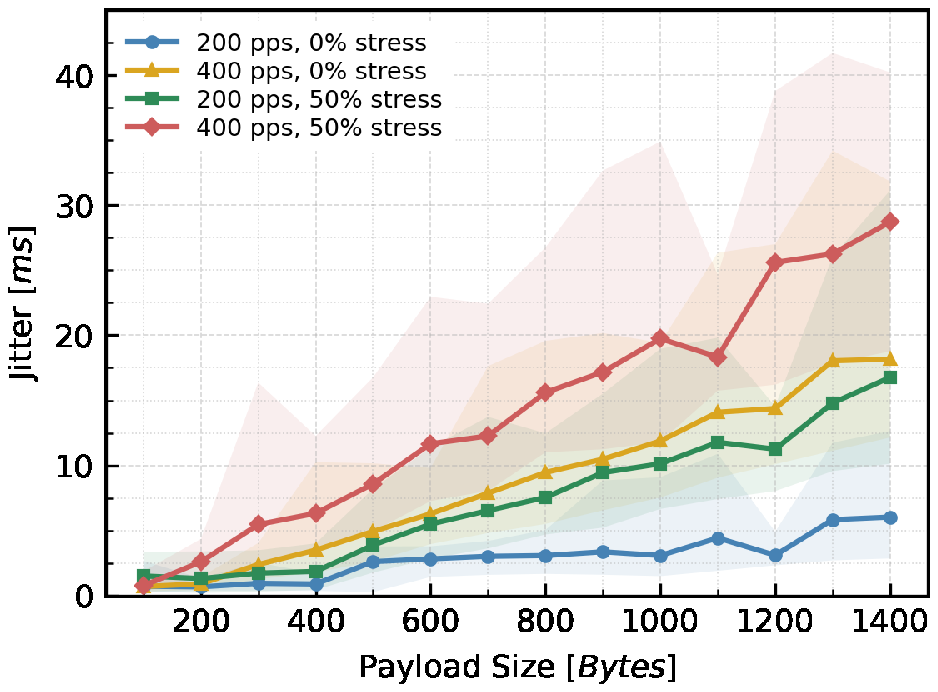}
    \subcaption{\textbf{Jitter \textit{(Enc)}}}
    \label{fig:jitter-enc}
  \end{minipage}
  \begin{minipage}{0.33\hsize}
    \centering
    \includegraphics[width=60mm]{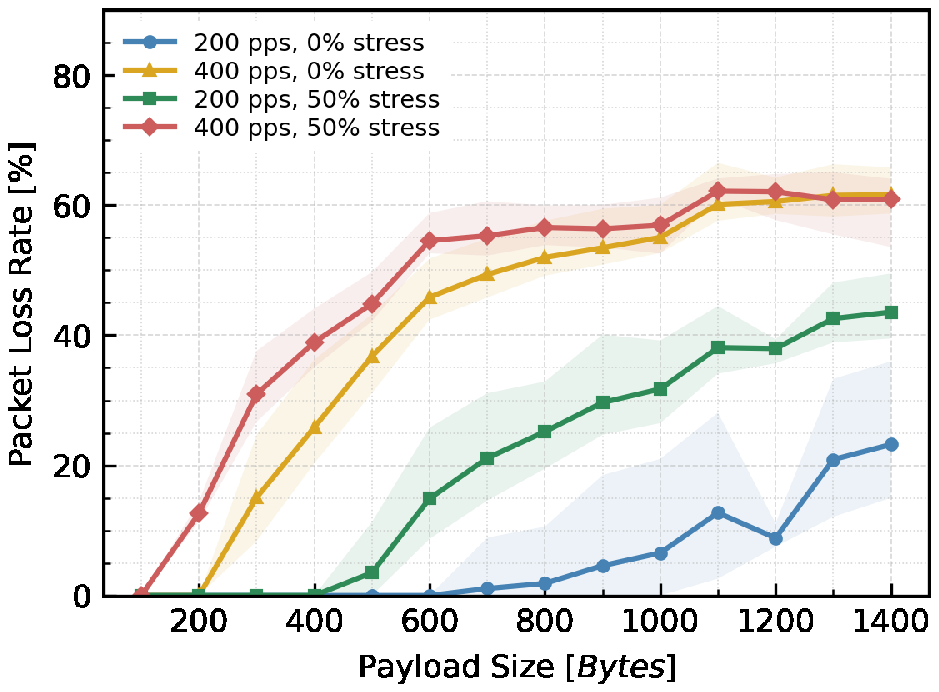}
    \subcaption{\textbf{Packet loss rate \textit{(Enc)}}}
    \label{fig:loss}
  \end{minipage}
  \label{fig:loss-jitter}
  \caption{\textbf{Jitter and packet loss rates:} 100th percentile values under several conditions.}
  \vspace{-2mm}
\end{figure*}

\begin{figure*}[!t]
  \begin{minipage}{0.33\hsize}
    \centering
    \includegraphics[width=60mm]{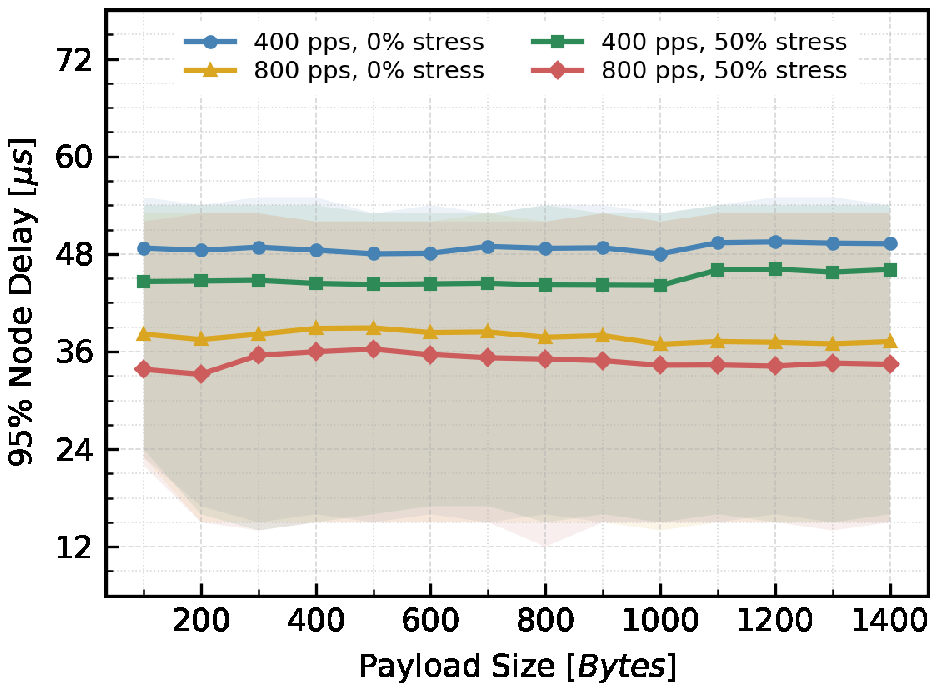}
    \subcaption{\textbf{Node delay \textit{(OLSR)}:} Only in-kernel delays}
    \label{fig:delay-olsr}
  \end{minipage}
  \begin{minipage}{0.33\hsize}
    \centering
    \includegraphics[width=60mm]{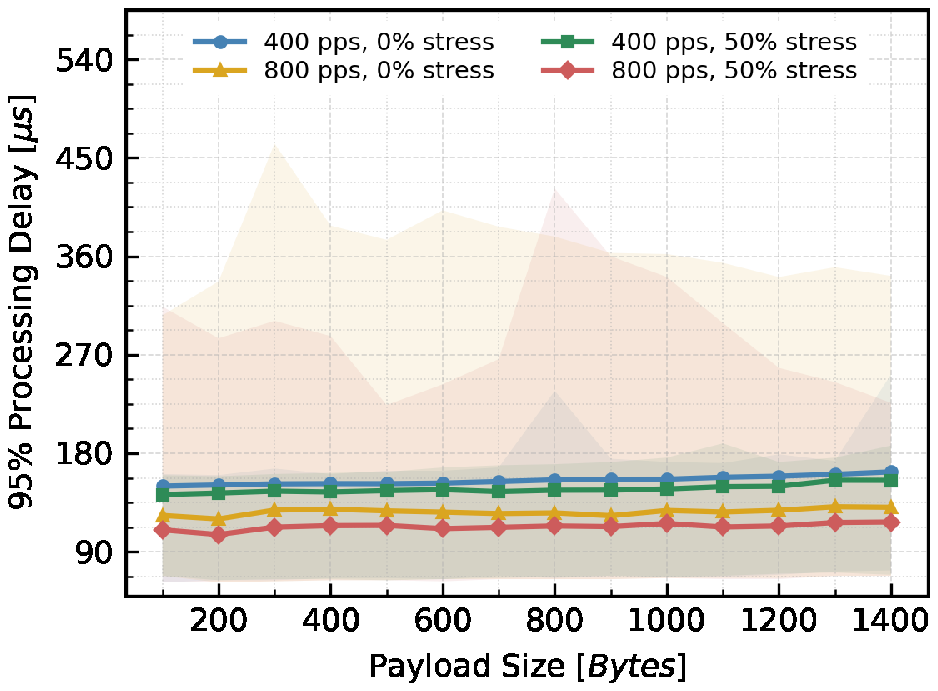}
    \subcaption{\textbf{\textit{Processing delay (AT)}}}
    \label{fig:delay-at}
  \end{minipage}
  \begin{minipage}{0.33\hsize}
    \centering
    \includegraphics[width=60mm]{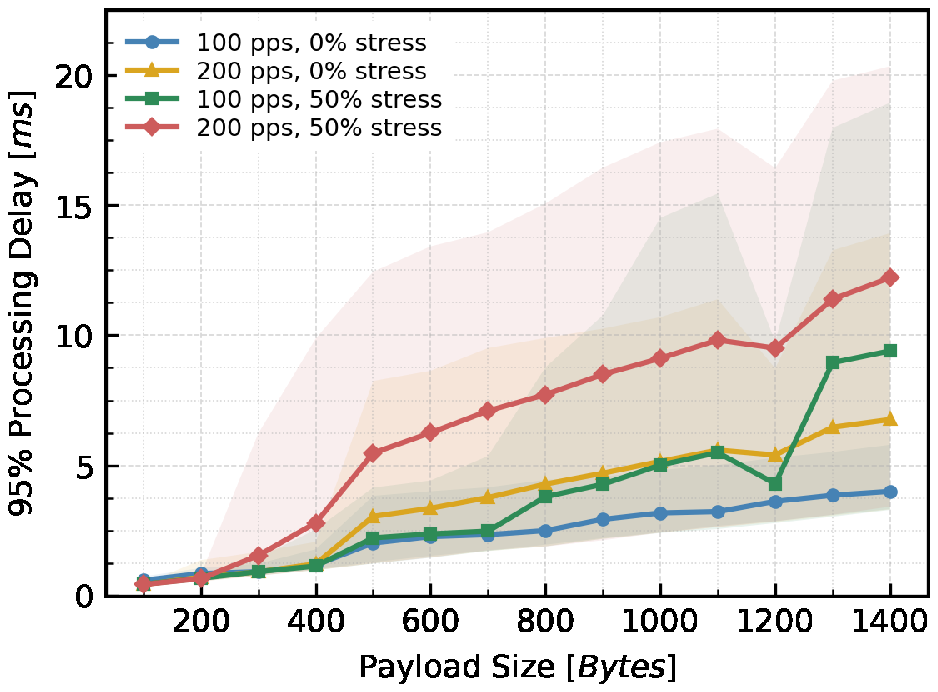}
    \subcaption{\textbf{\textit{Processing delay (Enc)}}}
    \label{fig:delay-enc}
  \end{minipage}
  \caption{\textbf{Node and \textit{processing delays}:} 95th percentile delays under the several conditions.}
  \label{fig:delay}
  \vspace{-2mm}
\end{figure*}

\begin{figure*}[!t]
  \begin{minipage}{0.33\hsize}
    \centering
    \includegraphics[width=60mm]{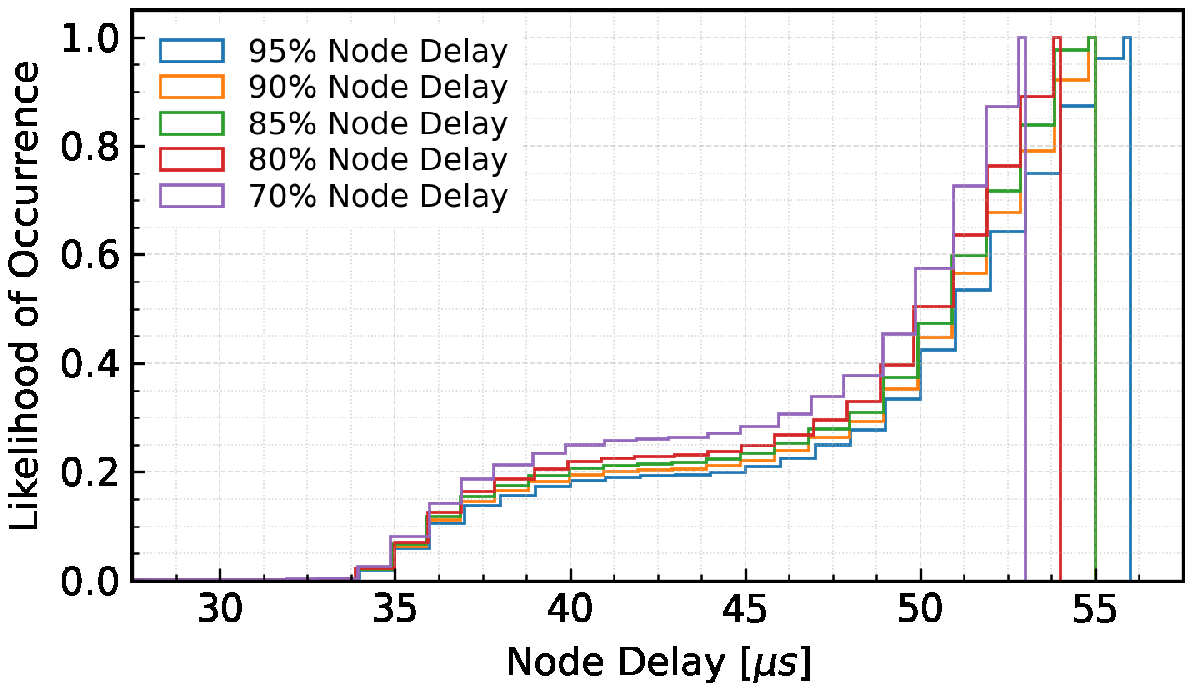}
    \subcaption{\textbf{Node delay \textit{(OLSR)}:} Only in-kernel delays}
    \label{fig:ecdf-olsr}
  \end{minipage}
  \begin{minipage}{0.33\hsize}
    \centering
    \includegraphics[width=60mm]{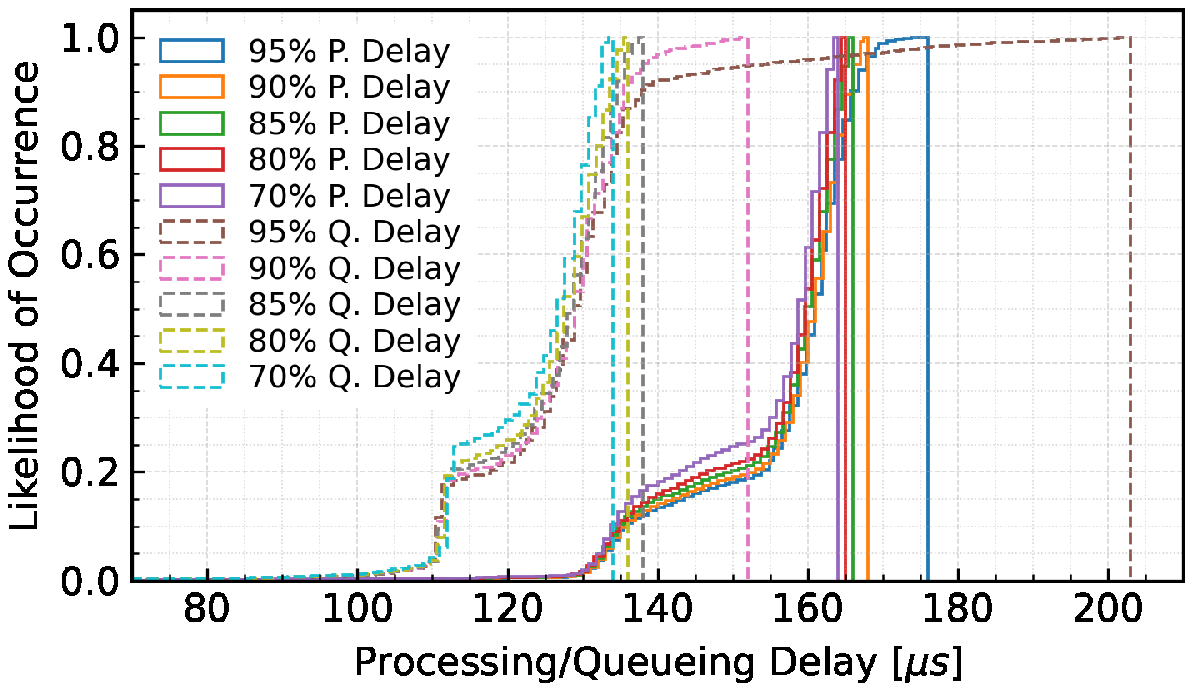}
    \subcaption{\textbf{\textit{Processing and queueing delays (AT)}}}
    \label{fig:ecdf-at}
  \end{minipage}
  \begin{minipage}{0.33\hsize}
    \centering
    \includegraphics[width=60mm]{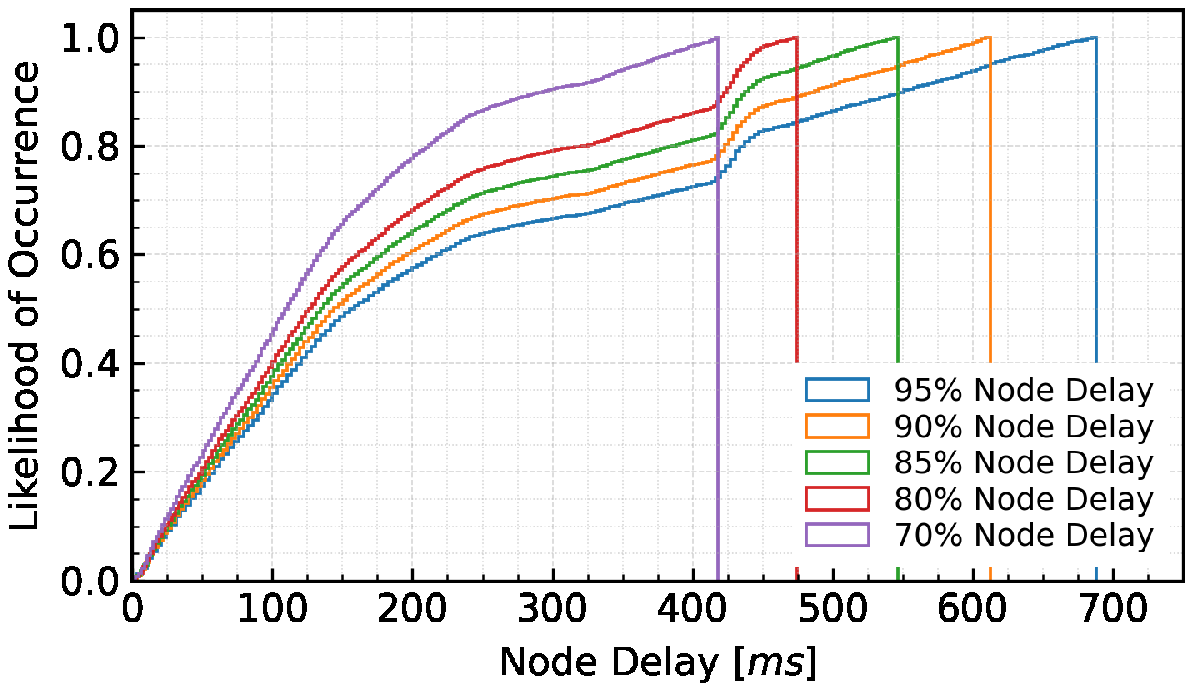}
    \subcaption{\textbf{Node delay \textit{(Enc)}}}
    \label{fig:ecdf-enc}
  \end{minipage}
  \caption{\textbf{Empirical Cumulative Distribution Function:} Several percentile node delays under 1000 bytes, 200 pps and 0\% \textit{stress}.}
  \label{fig:ecdf}
  \vspace{-3mm}
\end{figure*}

\begin{table}[!b]
  \begin{center}
    \small
    \caption{Parameter settings for measurement experiments.}
    \label{tbl:params}
    \begin{tabularx}{85mm}{ll}
      \hline
      Payload size & 100, 200, 300, ..., 1400 [bytes]\\
      Packets per second & 200, 400, 600, ..., 1200 [pps] \\
      Packets per second (\textit{Enc}) & 20, 40, 60, ..., 400 [pps] \\
      Additional CPU load (\textit{stress}) & 0, 50, 90 [\%]\\
      Transmission duration & 5 [sec]\\
      Number of samples & 50 \\
      \hline
    \end{tabularx}
  \end{center}
\end{table}

\subsection{Jitter and Packet Loss Rate}
Figures~\ref{fig:jitter-at} and \ref{fig:jitter-enc} shows the jitter of one-way communication delay.
Lines represent the average values, and we filled in the areas between the minimum and the maximum.
There were no significant differences between \textit{OLSR} and \textit{AT},
which suggests that lifting packets to the application layer does not affect jitter.
Jitter increased in proportion to the payload size only in the case of \textit{Enc}.

Similarly, only in the case of \textit{Enc} with 200 pps or more,
the packet loss rate tended to increase with payload size, drawing a logarithmic curve as seen in Fig.~\ref{fig:loss};
in all other cases, no packet loss occurred regardless of the conditions.

\vspace{-1mm}
\subsection{Node Delay}
Figure~\ref{fig:delay} shows the tendency of the node delay variation against several conditions,
and Fig.~\ref{fig:ecdf} shows the likelihood of occurrence as empirical CDF.

According to these figures, in the cases of \textit{OLSR} and \textit{AT}, the delay was nearly constant irrespective of pps and \textit{stress}.
There was a correlation between the variation and pps in \textit{OLSR}, while in \textit{AT} there was not;
this suggests that the application-level packet forwarding is less stable than kernel routing from the perspective of node delay.
In the case of \textit{Enc}, the \textit{processing delay} increased to the millisecond order
and increased approximately linearly with respect to the payload size, and the delay variance became large overall.
In addition, the graph tended to be smoothed as the pps increased;
this arises from the fact that packet encryption takes up more CPU time, which makes the influence of other processes less conspicuous.

It appears that the higher the pps, the lower the average delay (Fig.~\ref{fig:delay-olsr} and \ref{fig:delay-at}),
and the delay variance decreases around 1200 bytes (Fig.~\ref{fig:delay-enc}),
but the causes of these remain unknown, and further investigation is required.
One thing is certain: on the Raspberry Pi, pulling the packets up to the application through the network stack
results in a delay of more than 100 microseconds.

\vspace{-1mm}
\subsection{Link Delay vs. Node Delay}
\begin{figure*}[!t]
  \begin{minipage}{0.33\hsize}
    \centering
    \includegraphics[width=60mm]{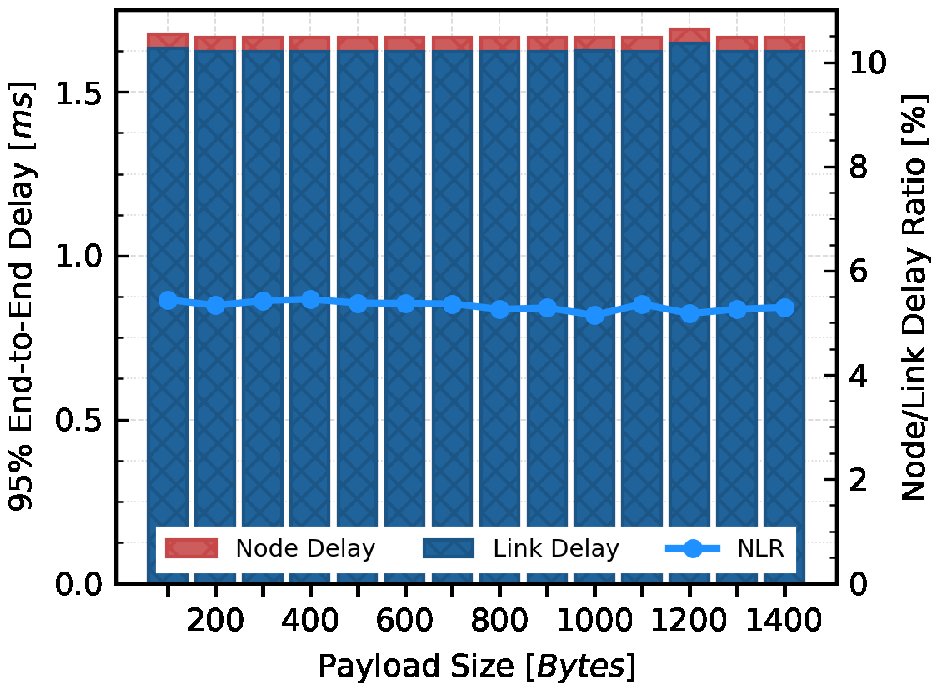}
    \subcaption{\textbf{E2E delay \textit{(OLSR)}:} 400 pps, 0\% \textit{stress}}
    \label{fig:total-olsr}
  \end{minipage}
  \begin{minipage}{0.33\hsize}
    \centering
    \includegraphics[width=60mm]{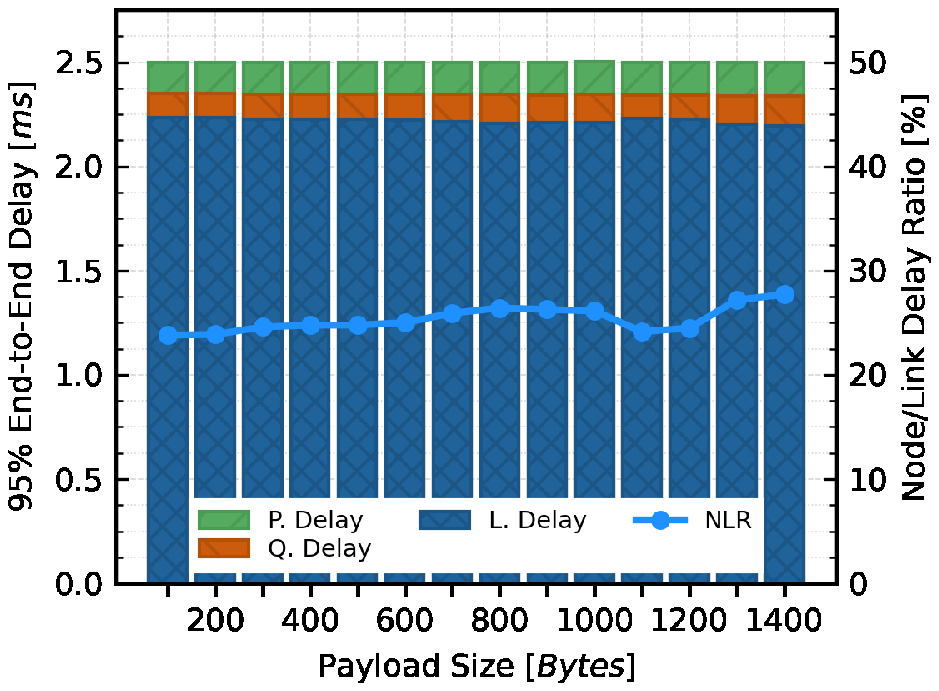}
    \subcaption{\textbf{E2E delay \textit{(AT)}:} 400 pps, 0\% \textit{stress}}
    \label{fig:total-at}
  \end{minipage}
  \begin{minipage}{0.33\hsize}
    \centering
    \includegraphics[width=60mm]{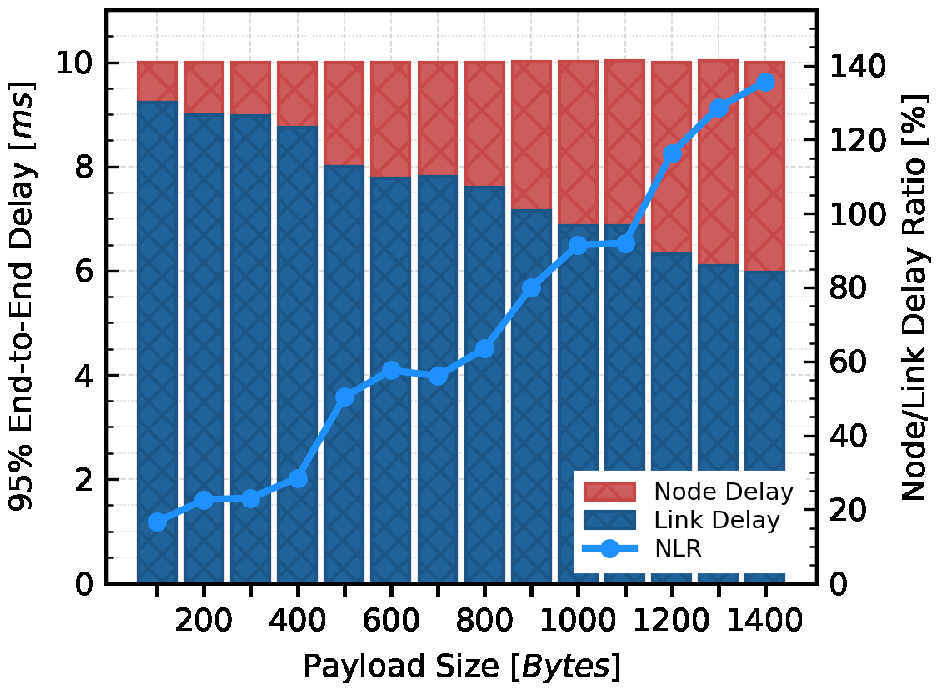}
    \subcaption{\textbf{E2E delay \textit{(Enc)}:} 100 pps, 0\% \textit{stress}}
    \label{fig:total-enc}
  \end{minipage}
  \caption{\textbf{End-to-end delay:} 95th percentile end-to-end delays including node and link delays.}
  \label{fig:total}
  \vspace{-5mm}
\end{figure*}

Figure~\ref{fig:total} shows the breakdown of the end-to-end delay and also describes the node delay link delay ratio (NLR).
As we saw in Fig.~\ref{fig:ad-hoc}, for this experimental environment, the end-to-end delay included two link delays,
and the link delay shown in Fig.~\ref{fig:total} is the sum of them.
The link delay was calculated from the effective throughput reported in iperf.
As iperf does not support pps as its option,
we achieved it by adjusting the amount of transmitted traffic, as

\vspace{-2mm}
\begin{equation}
  bandwidth = pps \times size \times 8 .
  \label{eq:pps}
\end{equation}

The results showed that, in the cases of \textit{OLSR} and \textit{AT}, the NLR was almost constant with respect to the payload size,
while in \textit{Enc}, it showed an approximately linear increase.
The NLR was less than 5\% in \textit{OLSR}, while in \textit{AT}, it was around 30\%, which cannot be considered negligible.
Furthermore, node delay was greater than link delay when the payload size was over 1200 bytes in \textit{Enc}.

\section{Conclusion and Future Work} \label{sec:conclusion}
In this work, we have designed and conducted an experiment to measure the software processing delay caused by packets relaying.
The experimental environment is based on an OLSR ad-hoc network composed of Raspberry Pi Zero Ws.
The results were qualitatively explainable, and suggested that, in low-energy ad-hoc networks,
there are some situations where the processing delay cannot be ignored.
\begin{itemize}
  \item The relaying delay of kernel routing is usually negligible, but when it is handled by application,
  the delay can be more than ten times greater, however simple the task is.
  \item If an application performs CPU-intensive tasks such as encryption or full translation of protocol stacks,
  the delay increases according to the linear model and can be greater than the link's transmission delay.
\end{itemize}

For this reason, node-to-node load balancing considering the CPU performance or amount of passing traffic
could be extremely useful for achieving delay-guaranteed routing in ad-hoc networks.
Particularly in heterogeneous ad-hoc networks (HANETs), where each node's hardware specs are different from each other,
the accuracy of passing node selection would have a significant impact on the end-to-end delay.

As we did not take any noise countermeasures in this experiment, our future work will
involve similar measurements in an anechoic chamber to reduce the noise from external waves
and an investigation of the differences in results.

\end{document}